# Probing temperature with lattice matched HgTe/CdTe core/shell nanoparticles

A.M.P. Hussain, S.N. Sarangi and S. N. Sahu[#]

*Institute of Physics, Sachivalaya Marg, Bhubaneswar-751005 (India)*

**Studies of photoluminescence (PL) can provide fundamental insight into the optical properties of semiconductor nanoparticles (NPs) but the same is being limited by NP size distribution and low luminescence yield [1]. Inorganic semiconductor core/shell [2-9] structures have added advantage of durability [2], high luminescence quantum yield (LQY), 10-80% in the visible range [2-5, 8] and useful in biological labeling [6, 9-12], LED [13] and luminescence thermometry [14-16] application. However, luminescence thermometry requires narrow PL line width, intense luminescence and should change with temperature linearly and reversibly which are difficult to achieve from lattice mismatched (3.9%-10.6%) [4] core/shell structures. Here, we report a lattice matched, durable and very small (2.8nm) HgTe/CdTe core/shell NPs whose optical activities are shown to be in the UV range and have yielded very high luminescence. The reversible change of PL with temperature of the core/shell NPs is used as an optical probe temperature sensor with resolution 0.32% per degree Kelvin which deemed to be one of the efficient luminescence thermometers reported till date.**

The colloidal chalcogenide semiconductor nanoparticles (NPs) are best known for their efficient fluorescence (FL) or photoluminescence (PL) in the visible [1,17,18], near infrared [19] and infrared [20] regions that depend upon NP size and are largely limited by NP size, size distribution and recombination loses at the surface trap sites [1] due to large unsaturated surface dangling bonds. The size distribution can be minimized by an organic ligand capping [21]. The luminescence quantum yield (LQY) and durability against photo-oxidation [2] can be improved by coating the NPs with an inorganic shell to passivate both anionic and cationic defects and confine [22] both electrons and holes to the core such that charge leakage across the shell is prohibited to minimize recombination loses. The surface passivated core/shell structures [2-9, 23, 24] can achieve LQY in the range of 10-85%. However, the lattice mismatch between the core and the shell of the epitaxially grown zinc blend core/shell structures are known to induce compressive and tensile strains to the core and the shell respectively which would change the band gap [25] of the core and shell respectively and would tend to lower the luminescence efficiency [26-28]. In addition, increasing the shell thickness induces a compressive strain on the core which results in the luminescence blue shift [5]. Furthermore, core/shell NPs with highly curved surfaces with good crystalline core can tolerate larger strain due to its distribution and relaxation at large curved surface constituent atoms of the NPs. LQY of 80% with CdTe/ZnTe [29] core/shell having 14% lattice mismatch has been reported where as 6.3% lattice mismatched CdSe/ZnSe core/shell could also achieve LQY of 85% [4]. In addition, 3.9% and 10.6% lattice mismatched CdSe/CdS [2] and CdSe/ZnS [3] core/shells respectively showed much lower LQY. For type II [30] core/shell ZnTe/CdSe core/shell, where their lowest electron energy states are spatially separated, the LQY is less than 20% even for a lattice mismatch of 0.6%. Furthermore, core/shell CdSe/ZnSe [4] and graded core/shell CdSe/CdS:ZnS [27] structures with lattice mismatch of 6.3 and 3.6% respectively have yielded LQY of 85% and 84%. However, majority of the higher lattice mismatched core/shells showed lower LQY [2,3,6,11,14]. This implies that the LQY not only depend upon strain relaxation along with good crystalline core curved surfaces but also require minimal lattice mismatch and uniform but optimum shell thickness. The core/shell structures can, thus, find meaningful applications in LED [13] and biological labeling [9-12]. A more interesting application of such high LQY NPs is in the area of luminescence thermometry as a temperature/pressure sensor in which the luminescence can act as an optical probe for aerodynamic surface temperature mapping [31] in wind tunnels. The requirement for such a sensor is that the device should (i) emit high yield monochromatic luminescence (ii) the change of luminescence be linear with the change of temperature and (iii) should have repeatability and reversibility in the temperature range under investigation. A narrow size distribution or monodispersed, lattice matched core/shell nanocomposite with high curved surface and high luminescence yield would be appropriate for optical temperature sensor applications.

[#] Corresponding author, E-mail: sahu@iopb.res.in, Phone: +91-674-2301058 extn-226
Fax: +91-674-2300142

Here, we report chemical synthesis of highly stable, lattice matched and L-cysteine ethyl ester hydrochloride (LEEH) capped fairly uniform size HgTe/CdTe core/shell nanocomposite with core size 2.0nm and shell thickness 0.8nm. HgTe and CdTe belong to the II-VI group of semiconductors with high absorption coefficients and band gap energies are respectively -0.15eV [32] and 1.46eV [33]. Strong excitonic absorption and FL occurred at 293.0nm and 370.0nm respectively have been observed from these core/shell nanocomposites. The reversible linear change of PL with temperature is used as an optical probe temperature sensor with resolution 0.32% per deg.K in the temperature range of 80-370 $^{0}$K.

The grazing angle x-ray diffraction (GXRD) of HgTe core and HgTe/CdTe core/shell are shown in figure 1(a) and 1(c) respectively which exhibit fcc structures with prominent <111> reflection and the peak positions do not differ from one another due to their similar lattice parameters (d=3.7Å). The GXRD peak intensities of the core/shell have been considerably decreased compared to the core [5] and the peak positions of both the core and core/shell have been shifted to higher reflection angles. The GXRD peak shifts and reduction in peak intensity of the core/shell is ascribed to the decrease of interplanar separation of crystal lattice and generation of compressive strain (table 1) in the samples due to the large surface to volume ratio and LEEH capping. The compressive strain, 0.25% of the core HgTe has been increased to 3.32% (table 1) for the HgTe/CdTe core/shell presumably due to large surface to volume ratio and LEEH capping. Figure 1(b) show the high resolution transmission electron microscope (HRTEM) images of core HgTe. The continuous lattice fringes across the crystal suggests good crystallinity of the core. The average size of the core HgTe has been estimated to be 2.0nm (from histogram) [21].

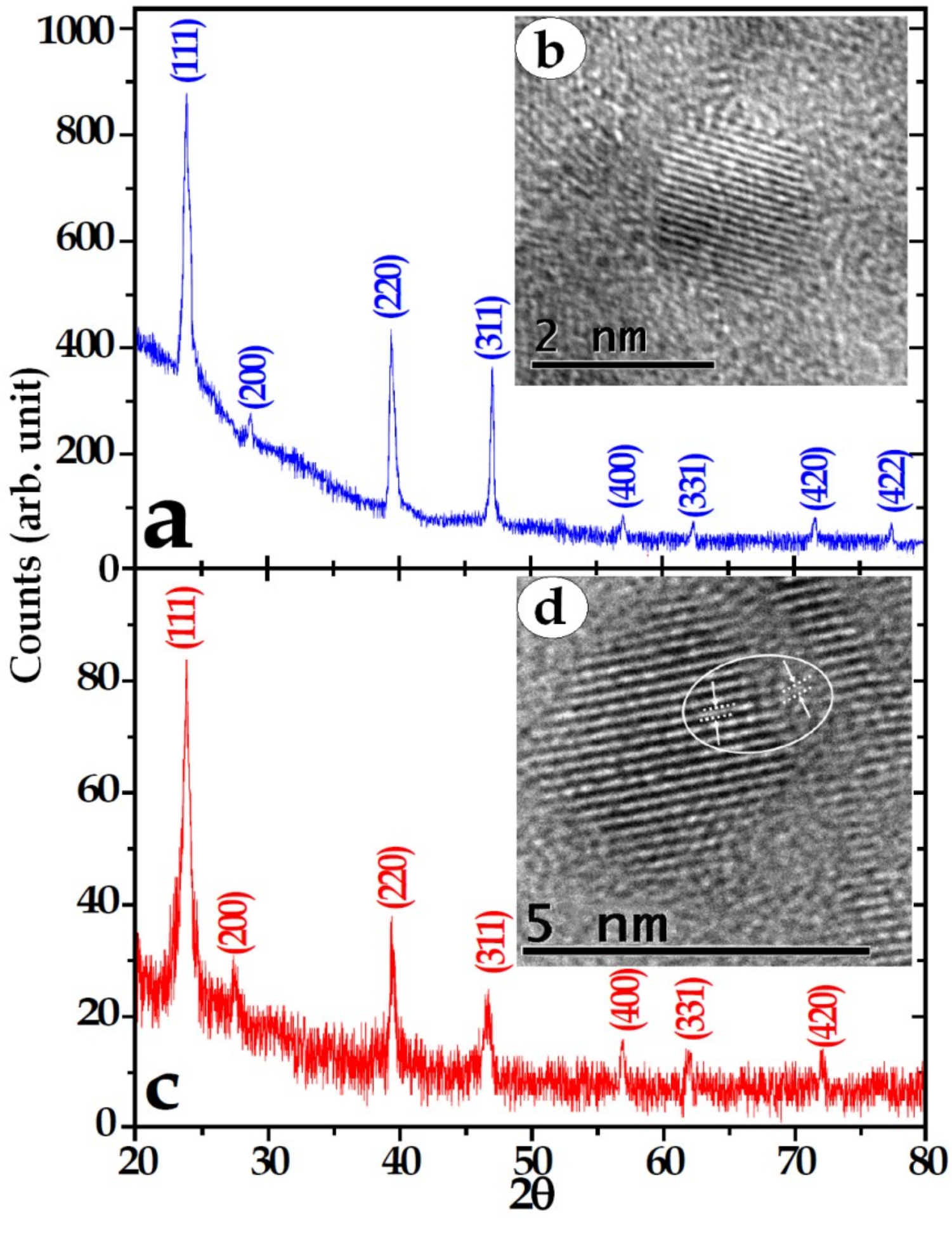

Figure 1: GXRD pattern and HRTEM micrograph of HgTe (a and b) and HgTe/CdTe core/shell NPs (c and d) respectively

Table 1: Average size of the HgTe NPs and HgTe/CdTe core/shell NPs estimated from UV-Vis, XRD and HRTEM studies and the compressive strain estimated from GXRD

| **Sample** | **Size (in nm) estimated by** | | | **Compressive strain (%)** |
|---|---|---|---|---|
| | **EMA** | **GXRD** | **TEM (average)** | |
| Core | 1.20 | 1.87 | 2.0 | 0.247 |
| Core/shell | 1.89 | 2.12 | 2.8 | 3.32 |

Figure 1(d) shows the HRTEM image of core/shell HgTe/CdTe with clear lattice fringes continuous across the entire crystal. The average size of the core/shell NPs obtained from the TEM micrograph and histogram (SI) is 2.8nm (table 1). The crystalline sizes estimated both from GXRD and TEM are comparable (table 1). An increase of size 0.8nm of the core/shell corresponds to a shell thickness [3] of 2.1 monolayer (mL) of CdTe shell which is consistent with those of earlier reports [3, 29]. The continuous lattice fringes across the entire core/shell structure suggest the epitaxial growth of CdTe shell over HgTe core. The bending of the lattice fringes marked by dotted line (Fig.1 (d)) exhibiting a well defined interface between the HgTe core and CdTe shell suggests a compressive strain at the interface [3] and is in agreement with the GXRD results. The increased compressive strain and reduction in GXRD peak intensities of the core/shell structure confirms the formation of a core/shell HgTe/CdTe rather than an alloying effect in which case the GXRD peaks in figure 1(c) should have been sharpened. The fcc structure of the core shifts to smaller bond lengths accompanied with the lattice constant shrinkage and results a quasi-spherical core/shell in agreement with earlier report [3] which suggests the growth of an uniform shell over the core despite a small compressive strain of 3.32% (table 1). Assuming that the entire amount of CdTe grows over HgTe core, the mean diameter of the core should increase by third root of the ratio of the molar concentration after and before coating. In the present case, the mean diameter of the core has been increased by 0.8nm nearly supports the earlier core/shell work [34].

The absorption spectra (Fig. 2, inset) of the fairly narrow size distributed HgTe core with average size 2.0nm [21] and HgTe/CdTe core/shell with average size 2.8nm estimated from TEM and histogram (SI) exhibit strong excitonic absorption peaks respectively at 280nm and 293nm in the UV region in contrast to other II-VI group [2-5, 29] materials and over all spectral feature are unchanged. The appearance of excitonic absorption peaks in the UV region and their peak broadening are accounted respectively to the strong size quantization effect (as the Bohr exciton radius of HgTe is 40nm [21] and the synthesized NP size is only 2.0nm) and small NP size distribution. The long wavelength tailing in the absorption spectra is ascribed to the scattering dominated absorption in powdery samples generally observed in chemically synthesized semiconductor NPs [35].

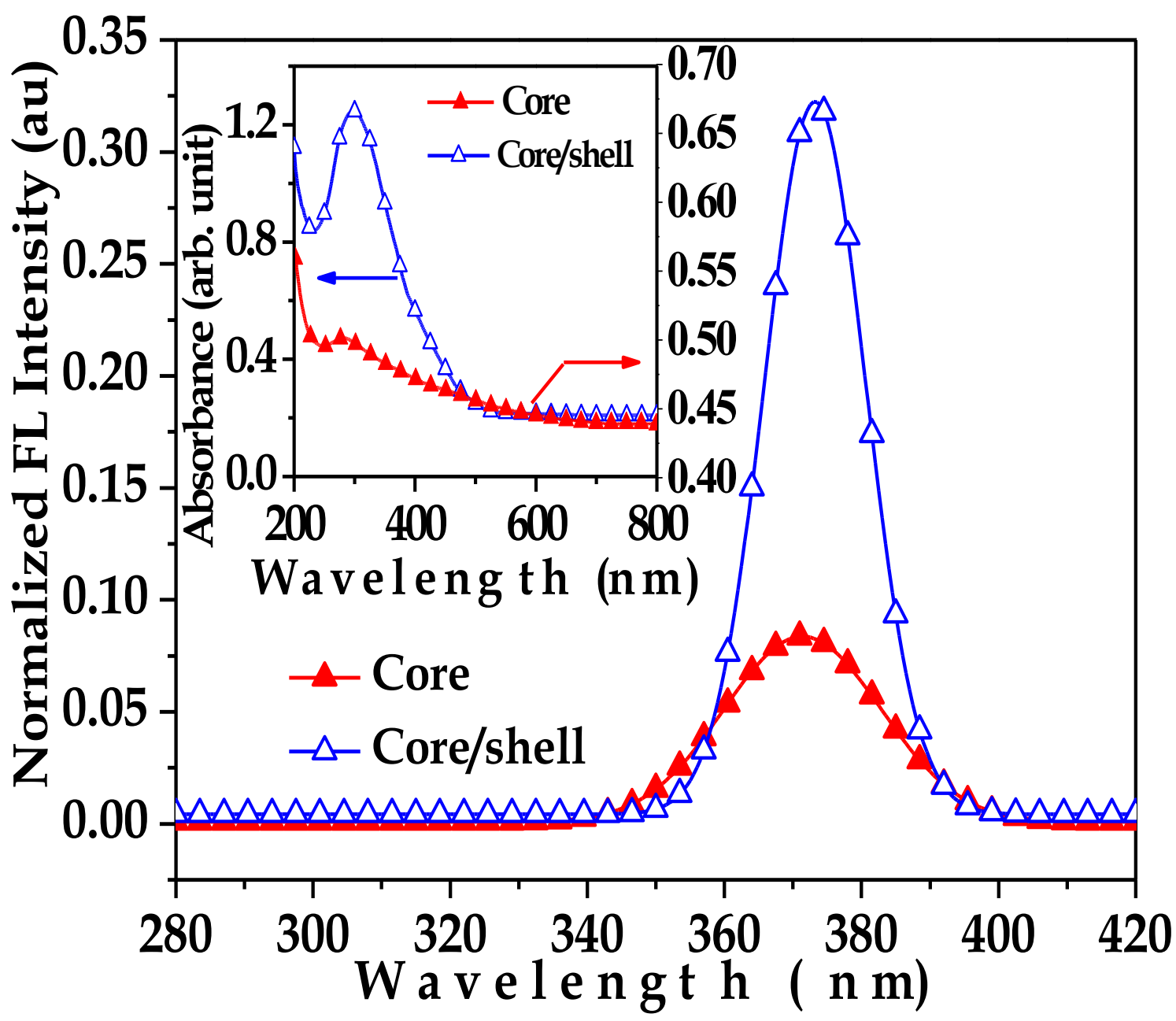

Figure 2: FL spectra of core HgTe and core/shell HgTe/CdTe nanoparticles with 250nm excitation, inset- UV-Vis absorption spectra of core HgTe and core/shell HgTe/CdTe nanoparticles.

A red shift of 13nm in the absorption peak of the core/shell with respect to the core is ascribed to the increase of core/shell size compared to the core which is also confirmed from GXRD and HRTEM studies. In addition, with the increase of the CdTe shell thickness, the exciton leaks to the shell and the peak should broaden due to the inorganic passivation of the core. Under this condition, the electrons leak to the shell and the hole stays in the core due to its large mass. The separation between electron and hole lowers the exciton confinement energy which ultimately results a red shift in the core/shell with respect to core. The crystalline sizes of the core and core/shell NPs were cross checked from optical absorption measurements using effective mass approximation (EMA) [36] and found to be less compared to those obtained from GXRD and HRTEM (table1). The discrepancy is ascribed to the non-sphericity of the NPs as well as limitation of EMA in the smaller size regime [37,38].

The FL spectra of the HgTe and HgTe/CdTe core/shells at 300 K under 250nm excitation wavelength are shown in figure 2. The FL of the core with narrow line width (= 18nm) and high colour purity appear at 370nm. Surface passivation of the HgTe core by the CdTe shell results progressive enhancement of FL yield, red shifted by 3nm in agreement with earlier reports [2-4,6,11,14]. The red shift of 3.0nm is due to (i) increase in core/shell particle size compared to core and (ii) compressive strain (table 1), although small.

In general, the cores of the II-VI core/shell NPs, already reported are either CdSe, CdTe or ZnTe [2-4,6,8,29] whose Bohr exciton radii are in the range of 4 to 6nms and the core/shell NPs sizes are in the range of 2-5nm. Hence, the quantum size effect (QSE) manifests in those core/shells which results in the overlap of first excitonic absorption peak with FL or PL and attributed to band edge recombination. However, in the present case, the Bohr exciton radius of HgTe is 40nm [21] and the synthesized HgTe NPs and the core/shell sizes are respectively 2.0nm and 2.8nm. Hence, very strong QSE should manifest in such core and core/shell structures where the electronic energy levels should highly be descritized which would lead to a large non-resonant stokes shift [39] which is inversely proportional to the size of the NPs [40]. The FL of the HgTe/CdTe core/shell does not overlap with the excitonic absorption peak, which suggests that the luminescence is not due to band edge recombination rather they are of excitonic nature. The large non-resonant stokes shift of 77nm is ascribed [39, 40] to a combined effect of (i) the fine structure of the band edge exciton, (ii) large coupling of the electron-hole pair in emitting state for LO phonons in the HgTe lattice for very small NPs (iii) formation of excimer within the very small clusters in concentrated solution and (iv) the organic capping [41]. Non-resonant Stokes shift as large as 150nm [35,42] in semiconductor NPs have already been reported. The lattice compressibility [43] of HgTe and CdTe are respectively 47.6 GPa and 44.5GPa. CdTe is most compressible among all the II-VI compounds except HgTe [44]. Further, CdTe has higher deformation potential [44] which implies that CdTe is softer and HgTe is harder. Over coating of HgTe by CdTe shell of 2.1 mL thick induces a small (3.25%) compressive strain at the core/shell interface. The core being very small (2.0nm) with highly curved surfaces, the strain instead of relaxing to generate trap sites, now distributed over the constituent atoms of the curved surfaces. The luminescence yield as high as 3.8 times to that of the core with high color purity in the present HgTe/CdTe core/shell NPs is being observed for an optimum shell thickness of 2.1 mL. Further, we have found that increasing the CdTe shell thickness beyond 2.1mL leads to spectral red shift with lower luminescence yield due to increasing compressive strain. Shell thickness less than 2.1mL again lowers the luminescence yield without much change in spectral position suggests to the non-uniform surface passivation. The line width (= 18nm) of HgTe/CdTe core/shell is much less than those of core/shells reported [2-4,6,11,14,29] earlier which suggest to the high color purity and narrow size distribution of the core/shell NPs. The core/shell NPs are highly stable against photooxidation as confirmed from the exposure of the NPs to UV light for 2 hours. We observed no change in the FL shape, width or position. Our experimental observation lead to the conclusion that to achieve intense luminescence from different choice of core/shell materials (i) the lattice mismatch between the core and the shell should be minimum, (ii) the core should be small with large curved surface, (iii) good crystallinity of the core and (iv) optimum shell thickness with uniform surface passivation.

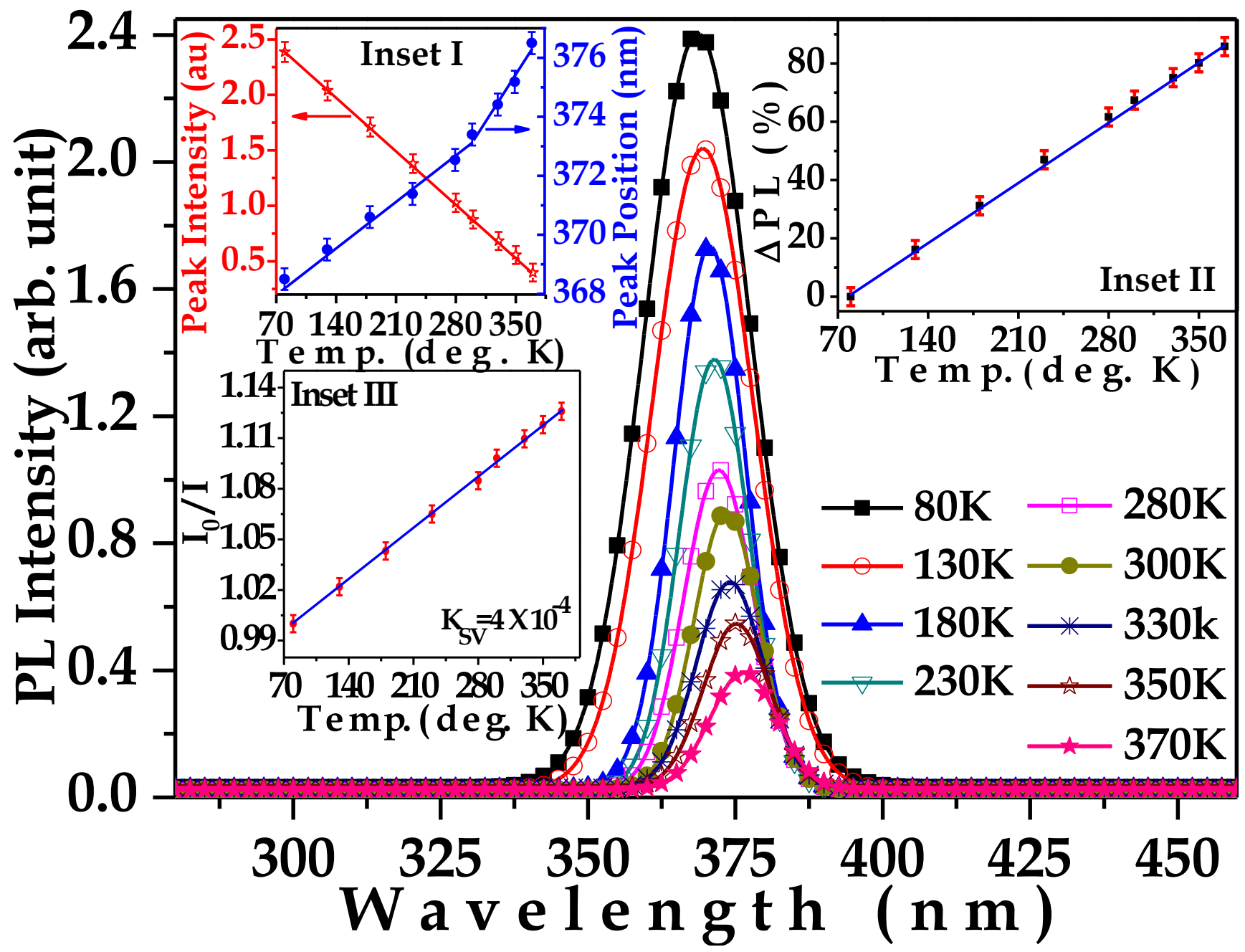

Figure 3: Temperature dependent PL spectra of HgTe/CdTe core/shell NPs, inset-I: PL peak intensity and peak position with temperature, inset-II: change in PL intensity as function of temperature and inset-III:- Stern-Volmer plot of PL intensity as a function of temperature where $I_0$ is the PL intensity at 80K and I is the PL intensity at respective temperature

The HgTe/CdTe core/shell NPs are stable in the ambient for more than five months and exhibited reversible PL intensity changes and small peak position shifts in the temperature range of 80 to 370K and provide the best possibility of luminescence thermometry applications such as aerodynamic surface temperature mapping [16] in wind tunnels. Hence, for such a temperature sensor application, the PL of the core/shell in thin film form on a glass substrate was probed in the temperature cycle 80k to 370K. The temperature was cycled from low to high and then back to low several times to test the reversibility of PL yield. The excitation wavelength was 250nm and PL was recorded at steady state condition at each temperature. The evolution of PL spectra as a function of temperature is shown in figure 3. Note that, two and half times increase of PL intensity at 80K as compared to 300K is observed due to the suppression of phonon vibration. Here, the ambient temperature is 300K. Below 300K, the PL is blue shifted and a maximum shift of 8nm occur at 80K. Above 300K, PL is red shifted and a maximum red shift of about 2nm occurs at 370K. Overall, the PL is red shifted with increasing temperature due to thermal expansion of lattice leading to band gap narrowing [45]. In addition, with increasing temperature, the PL is consistently quenched. The quenching of PL at high temperature is ascribed to the escape of carriers from the nanocrystals to the non-radiative recombination centers [46]. The PL position and intensity as a function of temperature is shown in figure 3(inset I) exhibit a linear trend. In terms of rate processes competing for deactivation of the lower excited singlet state under given conditions of temperature and dielectric, the PL yield can be expressed [47] as $\phi_y = k_f / k_f + \Sigma k_d$, where $k_f$ is the rate constant of PL and $\Sigma k_d$ is the sum of the rate constants for deactivation of the lowest excited singlet state by all competitive non-radiative processes. From the relation, as $\Sigma k_d$ increases, $\phi_y$ will decrease with increase of temperature. Indeed, with the increase of temperature, the deep traps became more active and contribute to the non-radiative processes. Unlike the linear decrease of PL intensity (figure 3 inset I) with increasing temperature, the red shift of PL peaks with increasing temperature is not linear. The PL shift generally depend upon inter dipole-dipole interaction [47] which possibly is not the same for all the temperature under investigation. As the core/shell was intended to be used as a temperature sensor, a calibration curve was plotted with change of PL with increasing temperature with the 80K as the reference. Such a calibration plot is shown in figure 3(inset II) in the temperature range of 80K to 370K. Linearity in the sensor calibration could be attained with a sensor resolution of -0.312% per deg.K which is slightly less than the reported result [14] whose temperature range is only 315K to 100K. Even the change in PL peak position with temperature is less than the reported results [14]. Such a sensor resolution with wide temperature range suggests that lattice matched and highly stable HgTe/CdTe core/shell NPs can be used as one of best temperature sensors. Figure 3(inset III) shows the Stern-Volmer plot [48] with linearity in the temperature range of 80K to 370K. The linearity and $K_{SV}$= 4 X $10^{-4}$ suggest dynamic quenching of PL with temperature.

In conclusion, we have demonstrated the fabrication of highly stable narrow size distributed and lattice matched HgTe/CdTe core/shells whose luminescence yield at 300K is much higher than any of the lattice mismatched strained core/shells. The colour purity of such core/shell structures is excellent. A clear core/shell interface can easily be identified and the strain relaxation at the curved surface of the shell without affecting the core has been proposed. The FL shows excitonic recombination associated with large Stokes shift. As the core/shells are photo stable and highly stable in the ambient, a temperature sensor have been fabricated with relatively high resolution (-0.32% per deg.K) in the wider temperature range, 80K to 370K. The possible and immediate application of such core/shell structures lies with multiplex biolabeling because of the hydrophilic LEEH capping, luminescence thermometry and electroluminescence devices.

**Methods:**

The LEEH capped HgTe/CdTe core/shell nanoparticles have successfully been synthesized in a simple two step chemical reaction. The precursor’s mercury chloride ($HgCl_2$), cadmium sulfate ($CdSO_4$) and tellurium powder (Te), reducing agent sodium borohydride ($NaBH_4$) and the capping materials LEEH, all 99.9% pure, are used as received. In all the steps, double distilled deionized water is used as the solvent. LEEH, a derivative of protein amino acid that exists naturally as a protein in most living organisms, is a hydrophilic surface passivating agent and do not show any absorption or emission in the UV-Vis region of our interest. The synthesis of HgTe is carried out as follows: 0.32mM of tellurium powder was reduced by 1mM sodium borohydride in water which produces NaHTe. To this, a solution of LEEH capped Hg (0.32mM) solution was added and stirred for 4 hours at $90^0$C to synthesize LEEH capped HgTe nanoparticles. The amount of capping agent LEEH was varied to 10, 15 and 20 times of the molar concentration Hg (3.2, 4.8 and 6.4 mM) to study the effect of capping agent on size and optical properties of the nanocomposites. In the second step, the formation of the shell was carried out to a portion of the LEEH capped HgTe as follows: 0.16mM solution of reduced tellurium powder (NaHTe) was added to water dispersed HgTe and stirred for 2 hours to make the surface of HgTe NPs Te rich to which stock solution of cadmium (0.16mM) was added for the formation of CdTe shell over the HgTe core. The CdTe shell was grown epitaxially over the HgTe core without conventional substitution reaction with either anion or cation. For temperature dependent PL measurements, a thin film of the core/shell nanocomposite is prepared by drop casting of NP dispersion on a cleaned glass substrate and dried under ambient. The X-ray diffraction of the NPs is carried out with a Bruker Advanced D8 GXRD. A 200 kV JEOL-2010 HRTEM fitted with a cooled CCD camera (Gatan) was used for the electron microscopy studies. The UV-Vis spectra of the NPs were recorded with a Shimadzu UV-3101PC spectrophotometer. The FL and PL behavior of the NPs in water dispersed state and in thin film form are recorded at room as well as controlled temperature using Oriel PL set-up with a PMT detector. The capping of the HgTe/CdTe core/shells by LEEH is confirmed from FTIR measurements (supporting information).

The LEEH capped HgTe/CdTe core/shell NPs are stable in the ambient even after 5 months as evidenced from their unchanged spectral shape and width in their absorption and fluorescence measurements. The photo stability test of the HgTe/CdTe core/shell NPs are carried out under 200mw/cm$^2$ intensity of illumination from a Hg-Xe lamp using a water filter to avoid the heating effect. The absorption and FL measurements at 300K exhibited no noticeable change after the continuous light exposure to the core/shell structures which establishes the high photo stability of the materials.

**Acknowledgement:** Authors would like to thank Prof. P.V. Satyam and A. Rath for their help in the HRTEM image recording.